\def\m{\multicolumn}
\documentstyle[]{laa}
\begin{document}
\thesaurus{08.1.3;08.05.2;08.08.1;11.13.1;11.19.4}
\title{Young LMC star clusters as a test for stellar evolutionary models}
\author{~A.~Subramaniam
        \and ~R.~Sagar }
        \offprints{~A.~Subramaniam}
\institute{Indian Institute of Astrophysics, Bangalore 560034 India}
\date{Received date: accepted date}
\maketitle
\begin{abstract}
We compare the observed colour-magnitude diagrams (CMD) and
the main sequence luminosity functions (LFs) of four Large
Magellanic Cloud (LMC) star clusters namely, NGC 1711, NGC 2004,
NGC 2164 and NGC 2214 with the synthetic ones derived from the stellar
evolutionary
models. Four different types of stellar
evolutionary models have been used for comparison.
The comparison of the features of the observed CMDs with the synthetic ones
favour the overshoot models from Bressan et al. (1993).
The synthetic integrated luminosity
functions from the models can be matched with  the observed ones
by varying the value of mass function slopes.  In order to constrain
the models from the comparison of the synthetic ILFs with the
observed ones, reliable estimates of mass function slope and
binary fraction are desired.
\keywords{LMC clusters - NGC 1711 - NGC 2004 - NGC 2164 - NGC 2214 -
theoretical stellar evolutionary models}
\end {abstract}
\section{Introduction} Colour magnitude diagrams (CMDs) of
young Magellanic Cloud (MC) star clusters are a very good testing ground
for the stellar evolutionary models (SEMs) in order to discriminate
among the possible
evolutionary scenarios of high mass to intermediate mass stars.
 Since clusters in the MC are rich, they
populate almost all the evolutionary phases and also
occupy regions of
the age and metallicity domain which are not populated in our
galaxy. They therefore extend the range of comparison between
SEMs and observational data. Furthermore, their study is
mandatory for the understanding of star clusters in external
galaxies where only integrated properties can be observed.

A detailed comparison of SEMs with the narrow and well defined stellar
sequences in the CMDs of MC star clusters could not be done
till recently, because most of the earlier CMDs (cf. Sagar \& Pandey
1989, Seggewis \& Richtler 1989) are based on either
photographic or electronographic observations which are,
in general, not only restricted to bright stars (V$\sim$17-18
mag) but also limited in accuracy. However, the advent of
modern detectors and software for doing accurate photometry
in crowded regions have recently made such work possible.
For example, Chiosi et al. (1989) analyzed the BV CCD data of
the cluster NGC 1866 to
disentangle classical stellar evolutionary models from
that incorporate convective core overshooting (CCO).  They
conclude that substantial overshooting ought to occur in stars
with mass of about 5 $M_\odot$.  Vallenari et al. (1991) also
support the above fact on the basis of an analysis of the CCD
data of NGC 2164. Bencivenni et al. (1991) while analyzing the
similar data for the cluster NGC 2004 conclude that their models
based on classical theory can very well explain the number of
stars present in the various evolutionary phases.  Stothers \&
Chin (1992) find that in the mass range 4-15 $M_\odot$, very
little or no overshooting of the core is needed to reproduce the
observed features of the blue populous clusters, NGC 330 and NGC
458 in the Small Magellanic Cloud. This leads to
a confusing state regarding the need for CCO in the models for various
masses and its amount.  The situation gets further complicated
by the
use of different opacity tables in different SEMs. Various
tables of radiative opacity have been published by the Livermore
group over the last few years, the two recent ones being
from Rogers \& Iglesias (1992, hereafter referred to as OPAL1)
and from Iglesias et al. (1992, hereafter referred to
as OPAL2).  These modify the surface parameters of the stars
significantly in comparison with the opacities from Los Alamos
Opacity Library (LAOL) of Huebner et al. (1977).
Stothers \& Chin (1991) show that in the structure calculations
of SEMs, a large assumed metal abundance can remove the need for
assuming significant CCO. At the same time Schaller et al.
(1992) and Bressan et al. (1993) show that such an assumption
has little effect on the lifetimes and lifetime ratios of the
central H and He-burning phases.
 Here we
try to address the above questions by comparing synthetic CMDs and
luminosity functions (LFs)
obtained from the SEMs given by Castellani et al. (1990),
Schaller et al.  (1992) and Bressan et al. (1993) with the
observed CMDs and LFs of the four LMC star clusters namely, NGC 1711,
NGC 2004, NGC 2164 \& NGC 2214.
The four clusters considered span a wide
range in age, thus have a range in turn-off masses. This
analysis will be very helpful in discriminating the properties
of these different masses as predicted in the models.  The
remaining sections contain the method adopted for incompleteness
correction and field star subtraction, a brief description of
the SEMs and the determination of the age of the clusters. The
results of comparison between the predicted and the observed
CMDs and LFs and conclusions follow.
\section{Observational data}
 The data for the clusters NGC 1711, NGC 2004, NGC 2164 \& NGC
2214 are taken from Sagar et al. (1991b) (hereafter referred to as Paper I).
The data were obtained at the f/8.5 Cassegrain focus of the
1.54m Danish telescope at the European Southern Observatory, La
Silla, Chile using a RCA CCD chip of 320$\times$512 pixel in
size where one pixel corresponds to 0.47 arcsec. Other details
of observations and data reductions are given in Paper I. Since
the data for
the field region of NGC 2164 in Paper I is not very deep, we
have taken those from Vallenari et al. (1991).

Using secondary photoelectric standards covering a range in
brightness (7.3$\le$V$\le$12.4) and in colour
($-$0.17$\le$(B$-$V)$\le$1.68), the photometric data have been
calibrated with a zero-point accuracy of $\sim$ 0.04 mag in both
B and V. The present CCD data agree well with independent
photoelectric observations of some stars in the clusters having
a large range in brightness (11.8$\le$V$\le$18.1) as well as in
colour ($-$0.17$\le$(B$-$V)$\le$2.17) but generally show
systematically varying differences with photographic data (cf.
Paper I). In the case of NGC 2004 \& NGC 2164 a comparison of the
present CCD data with independent CCD observations made in Paper
I indicate good agreement in the case of NGC 2164 but a
significant systematic difference in the case of NGC 2004.
Recently CCD observations generally for stars brighter than
V$\sim$17 have been published by Elson (1991) in the regions of
all our programme clusters; by Kubiak (1990) in the region of
NGC 1711, by Balona \& Jerzykiewicz (1993) for NGC 2004 and by
Bhatia \& Piotto (1994) for NGC 2214. The last two have
compared their data with ours and found a small systematic
difference, within the accuracy of zero-points, relative to
their measurement.  This indicates that these independent CCD
measurements agree within zero point errors of the photometry.
We plot the photometric differences with Kubiak (1990) and Elson
(1991) in Figure 1 and present the statistical results of the
comparison in Table 1.  Elson (1991) indicates that typical
uncertainties in her V measurements are $\pm$0.10 mag for the
fainter stars and $\pm$0.05 for the brighter stars, while those
in (B$-$V) are typically of 0.10 mag. However, the maximum
uncertainty in (B$-$V) could be up to $\pm$0.50 mag. According
to Kubiak (1990), his data can have a systematic error of $\sim$
0.10 mag in the zero point of (B$-$V) and slightly less in V.
Considering these facts, we conclude that our data agree fairly
well with Elson (1991) but have systematic differences with
Kubiak (1990). The differences in (B$-$V) are minimum
($\sim$$-$0.04 mag) for bluest stars but increase systematically
to become $\sim$0.5 mag for redder stars. A straight line fit
through the data points using least squares linear regression
gives, $$\Delta(B-V) = -0.30(\pm0.014)(B-V) - 0.10(\pm0.01)$$
with correlation coefficient, $\gamma$ = 0.98. This may indicate
improper use of colour coefficients in the photometric
calibration by Kubiak (1990) which yields only (B$-$V)$\sim$1.0
mag even for the reddest stars.

\section{Observational CMDs and LFs}
 The CMDs of the clusters analyzed here are presented in Paper
I. These CMDs need to be corrected for data incompleteness and
field star contamination before they can be compared with SEMs.
The first step towards this goal is to define a cluster region
which can be considered as cluster representative, but suffering
less from the above defects. The next task is to determine the
incompleteness of the data and adopt a suitable method to remove
the field stars from the CMDs.
\subsection{Selection of the cluster region}
 The idea is to find an annular ring best representing the
cluster but suffering less from incompleteness and crowding. The
cluster center is derived iteratively by calculating the average
x and y positions of stars within 150 pixels from an eye
estimated center, until it converges to a constant value. An
error of a few arcsec is expected in locating the cluster
center.  As the crowding is maximum near the center we expect
the data to be least complete near the central region.  We
consider the region where the data completeness factor (CF) is $<$ 30 \%
(defined in Sagar \& Richtler 1991; hereafter referred as Paper
II) as the central region and exclude this circular region from
our further analysis. This serves as the inner boundary for the
required cluster region. The outer boundary is identified from a plot of
stellar density as a function of the distance from the cluster
center. The limits of the annular region best representing the cluster
are given in Table
2. Since incompleteness varies also within this region, we divide
it into two rings, namely, ring 1 and ring 2 as given in
Table 2.

\subsection{Incompleteness correction}
The procedure to quantify the stellar completeness factor CF in
photometry is described in detail in Paper II.  The completeness
factor $\Lambda_i$ at a point ($V_i$, (B$-$V$_i$)) in V, (B$-$V)
diagram will be mainly controlled by that CF value of B and V
CCD frames where completeness is less, i.e., the value of
$\Lambda_i$ cannot be larger than the smaller value of the pair
$(CF(V_i), CF(B_i))$; where CF($V_i$) and CF($B_i$) are the
completeness values at the brightness $V_i$ and $B_i$ in the V
and B CCD frames respectively.  Consequently, we have used,
\begin{equation}
        CF(V_i,(B-V)_i) = min (CF(V_i), CF(B_i)),
\end{equation}
for the data incompleteness correction in our analysis. The
completeness values for all regions are taken from Paper II
except for the field region of NGC 2164, where they are from
Vallenari et al. (1991). The value of CF($B_i$) is generally less
than that of CF($V_i$) except in a few cases (see table 4 of
Paper II).

For applying the incompleteness correction to get a complete
CMD, the CMDs for the rings 1 and 2 are constructed
separately. Then the CMD of each ring is divided into a number
of boxes having a width of 0.2 mag in V and 0.1 mag in (B$-$V).
The $\Lambda_i$ for the mean values of $B_i$ and $V_i$ for
$i^{th}$ box is found using the relation (1).  The actual number
of stars in the $i^{th}$ box is then calculated as $N_{ci} =
(N_{oi})/{\Lambda_i}$, where $N_{oi}$ is the observed number of
stars in the $i^{th}$ box. Now, the extra number of stars in the
$i^{th}$ box, $\triangle N_i$, given by, $\triangle N_i = N_{ci}
- N_{oi}$, is then randomly distributed inside the box. In this
way the stellar incompleteness is corrected for both the rings
and then the CMD of the rings are combined to obtain a complete
CMD for the chosen cluster region.  The same technique is
applied to the field region without dividing it into different
rings since the incompleteness does not depend on position
here.
\subsection{Correction for field star contamination}
We have used the zapping
technique described by Mateo \& Hodge (1986) to remove
the field stars from the cluster CMD after accounting for the difference in
the areas of field and cluster regions considered.
In this method, for each star in the CMD of the field region, the
nearest star in the CMD of the cluster region
is identified and removed. The maximum box size for such an
identification is varied and finally fixed at $V \sim 1$ mag and
$ B-V
\sim 0.5$ mag. This box size is not related to the one used for the
completeness correction.
Lupton et al. (1989) have measured the radial velocities of some
stars in NGC 2164 and NGC 2214 and identified a few of them either as
galactic foreground stars or LMC field stars.
We have therefore removed them from our further analysis. In
this way, we get a cluster CMD ready to be compared with the
theoretical SEMs.

\section{Observed CMDs}
For all the programme clusters, Figure 2 shows the observed CMDs
corrected for photometric incompleteness and field star contamination.
For comparing the number of stars present in
different stellar evolutionary phases, the CMDs are divided into
three parts consisting of MS stars, bright red giants (BRGs) and
faint red giants (FRGs). The BRGs are the evolved stars of the
cluster and the FRGs are the intermediate age core helium burning
stars of the LMC field. Even after the field star subtraction
using zapping technique we do find some FRGs, which we reject
assuming them as the left out field stars.  Here we
discuss the various features of the final cluster CMDs and the
main sequence LFs.
\begin{description}
\item [(i)] NGC 1711
has a well populated MS ($\sim$590 stars) up to $V \sim 16$
and some stars are seen scattered up to 14.2 magnitude.
A total of 10 BRGs are distributed between 0.0 and 2.0 mag in (B$-$V). The
red clump of BRGs are at a (B$-$V) value of 1.6.
\item [(ii)]
NGC 2004 is the youngest of the clusters considered here,
having a well populated
MS of $\sim$520 stars continuous up to $V \sim 13.5$ mag.
There are six BRGs confined to a very small range in
brightness, 14.5$\le$V$\le$15.25 but with somewhat larger range
in their colours, 1.3$\le$(B$-$V)$\le$2.0.
\item [(iii)]
The brightest MS star in NGC 2164 has V$\sim$14.5 mag and
(B$-$V)$\sim$$-$0.1. Then there are no MS stars up to V$\sim$15
mag. Below that the cluster MS is well populated with $\sim$530
stars. The 16 BRGs are seen scattered between
$-$0.1 and 1.4 mag in (B$-$V).
\item [(iv)]
Recently, the cluster NGC 2214 has drawn a lot of attention
regarding its peculiar nature. This cluster is understood to
have two components (cf. Bhatia \& MacGillivray 1988), but the
debate is on the ages of these two components as to whether they
are equal or different. Sagar et al. (1991a) not only
assign two different ages for the two components but also
indicate that the older population is more centrally
concentrated (R$<$50 pixels) while
the younger one shows a more extended distribution. However,
recently, Lee (1992) and Bhatia \& Piotto (1994) have questioned
the reality of the presence of the older population. Since we
are not considering the central portion in our analysis here, we
are omitting the controversial older population.  The
cluster MS containing $\sim$550 stars is well populated up to
V$\sim$16 mag.  The V magnitude difference between the
brightest and the next brightest MS star is $\sim$0.8 mag.
The 14 BRGs are populated  between $-$0.1 and 1.3 mag in (B$-$V).
\end{description}

\section{Luminosity functions of the MS stars}
The differential luminosity functions (DLFs) for the MS stars
are calculated from the CMDs discussed in last section using a
bin width of 0.2 mag in V. The resulting DLFs for the
clusters under study are shown in Figure 3, where the effects of
correction of photometric incompleteness as well as of the field
star contamination are also indicated.  Despite the incompleteness correction,
the luminosity functions seem to be complete only for stars brighter
than V$\sim$19.5 mag. We have fitted a straight line to the log N vs
V plots using the least squares regression. The value of the slopes are
0.16$\pm$0.03, 0.19$\pm$0.04,
0.15$\pm$0.03 and 0.16$\pm$0.02 for the clusters NGC 1711, 2004,
2164 \& 2214 respectively. This indicates that the slopes of
their mass functions are similar which is in agreement with the
results given in the Paper II.
\section{Brief discussion of evolutionary models}
We use four SEMs in this analysis, two classical models and two
with CCO.
Both CCO models use the new opacity values. The Z value
considered is 0.02 for all the models.

The new opacities are significantly higher
than the LAOL around two temperatures, i.e. a few 10$^5$ K and
about 1 10$^6$ K. For solar metallicities, the difference in opacity
amounts to a factor of three and 20\% respectively (cf. Bressan
et al. 1993). The OPAL2 supersedes the previous OPAL1 because of
the inclusion of the spin-orbit interaction in the treatment of
Fe atomic data and the adoption of the recent measurements of
the solar photospheric Fe abundance by Grevesse (1991) and
Hannaford et al. (1992).

The differing input physics and
computational techniques from model to model makes a direct
comparison very difficult. The necessary
homogeneity is provided by the Bressan et al. (1993) models as
they include both classical and CCO models with all the other
inputs remaining the same.
The details
of the models are given below:
\begin{description}
\item[a] Model 1
given by Castellani et al. (1990) is a classical model. It
covers a mass range of 0.8 to 20$M_{\odot}$.
This model is used to
explain successfully the LF of NGC 2004 by Bencivenni et al.
(1991) and NGC 1866 by Brocato et al.  (1990). They have used
old opacity tables (LAOL) for log T > 4 and at low temperatures
opacities are taken from Cox \& Tabor (1976). The evolutionary
sequences are computed from zero-age main-sequence (ZAMS) to the
thermally pulsing asymptotic giant branch (TPAGB) phase.
\item[b] Model 2
given by Schaller et al. (1992), cover a mass range of
0.8 $M_\odot$ to 120 $M_\odot$. They have
used new opacities (OPAL1) supplemented by radiative opacities by
Kurucz (1991) at temperatures below 6000 K. As a result of
the increased value of opacities, the
convection parameters $l/H_p$ and the overshooting distance
change with respect to their previous model given by Maeder
\& Meynet (1989), where $l/H_p$ is the ratio of mixing length to
the pressure scale height. The value of $l/H_p$ is $1.6\pm0.1$.
The ratio of overshooting distance to the pressure
scale height, $d_{over}/H_p$ is 0.2 in the mass range 1.25 to
25 $M_\odot$.  Mass loss is included for all the stellar masses
throughout the HR diagram.
\item[c] Model 3
given by Bressan et al. (1993) present SEMs in the mass range
of 0.6 to 120 $M_\odot$. The evolutionary tracks extend from
ZAMS to TPAGB phase for low and intermediate mass stars, and
to the central carbon-ignition for higher masses.  The opacities
at temperatures between 6000 K and 10$^8$ K are taken from OPAL2.
The opacities at lower and higher temperatures are taken from
LAOL and Cox \&
Stewart (1970a,b) respectively. The value of $l/H_p$ is taken as 1.63.
The value of $d_{over}/H_p$ is
0.25 for the mass range 1.0 to 1.5 $M_\odot$ and equals 0.5 for masses
above. This value of 0.5 corresponds to the overshoot distance favoured by
Schaller et al. (1992). They have also included the effects of envelope
overshooting at the bottom of the convective envelopes, by
adopting a value of 0.7 for the envelope overshooting distance.
The mass loss is included only for stars more massive than 12 $M_\odot$.
\item[d] Model 4 given by Bressan et al. (1993) is same as model 3
without incorporating the core overshooting but with all
other inputs being the same. This set is available within the mass
range 2.5-20 $M_\odot$.
\end{description}

\section{Determination of age of the clusters} To determine the
age of the clusters, we fit the isochrones obtained from the SEMs
mentioned above to the ($M_V$, (B$-$V)$_O$) CMDs of the clusters
obtained from their V,(B$-$V) CMDs. We adopt a value of 18.6 for the
distance modulus to the LMC based on the recent studies on cepheid
variables in LMC (Viswanathan 1985, Caldwell \& Coulson 1985, Walker
1987, Welch et al 1987, 1991 ). The reddening values used by us are
given in the Table 2. For NGC 1711, the E(B$-$V) value is taken from
Mateo (1988) and for others from Cassatella et al. (1987). We
assume the extinction in V as 3.1 $\times$ E(B$-$V).
For fitting the theoretical isochrones in the M$_V$, (B$-$V)$_O$
diagram, we converted them from the log L/L$_\odot$ vs log T$_{eff}$ plane to
the  M$_V$, (B$-$V)$_O$ plane using colour-temperature
relations and bolometric corrections from Kurucz (1979) and
Vandenberg (1983) complemented with values given by Johnson (1966)
for the temperatures below 4000$^o$K.  The isochrones fitted to the CMDs of
the clusters under study are shown in Fig. 4. The ages and the
turn-off masses obtained from different models are given in Table 3
which indicate that they depend strongly on the SEMs used. Isochrones
with CCO make the object older but the turn-off mass lighter while
those without make the cluster younger but the turn-off mass
heavier. The ages and the turn-off masses of the clusters derived
from the two classical models agree very well with each other. This
may indicate that use of new opacities in SEMs has not changed
the hydrogen burning time scales significantly. Amongst the CCO
models, model 3 makes the object oldest, even though the
overshoot amount is the same as in model 2.
However, both the CCO models yield the same turn-off masses. We also
notice that the red giant stars are generally best fitted by the CCO
models. The isochrones produced from the classical models could not
reach the observed red end of the giant branch.  Bencivenni et al.
(1991) using the model 1 found an age of 8 Myr for the cluster NGC
2004, but using the same model we find an age of 12Myr. We find that
an age of 8 Myr predicts the BRGs roughly 1 mag brighter than the
observed ones, instead a 12 Myr isochrone fits the red giants better
though it falls short by $\sim$ 0.25 in (B$-$V) mag.

\section{Comparison of observed CMDs with synthetic CMDs}
In order to compare the theoretical stellar evolutionary models
with present observations, we produced synthetic CMDs from the
SEMs discussed in section 6 using the method described in the
Appendix. A comparison of the observed distribution of stars in
the CMD of a cluster with the that of synthetic CMD produced
from different SEMs tell us about the reliability of the
physical assumptions involved in theoretical calculations. For
this we compare the features as well as the integrated luminosity
function (ILF) of the observed CMDs with the synthetic ones.
\subsection{Comparison of features}
Here, we compare the features of the synthetic CMDs with the
observed ones. The synthetic CMDs for the clusters which are
best matching with the observation are shown in Fig. 5.
The absolute magnitude of the MS tip, the absolute magnitude of the
faintest BRG, the (B$-$V) value of the reddest BRG and the number of
blue supergiants predicted by the models and the values obtained
from the observed CMDs are tabulated in Table 4. Models 2 and 4
predict blue supergiants in all the clusters which is contrary to
the observation, while
models 1 and 3 produce almost none in their synthetic CMDs.
In all the clusters the evolved part of the
CMD is more closely reproduced by Model 3. It is interesting to
note that even though the amount of CCO is the same in models 2
and 3, the number of blue super giants and the evolved parts of
the CMDs are better matched with model 3. This may be due to the
differing input physics and computational techniques involved
in the two models.

In the cluster NGC 2004, the brighter end of the MS is not populated up to the
observed value by any of the
synthetic CMDs. Therefore we tried to populate the MS up to the
observed tip by adopting an age spread. This gives rise to a
spread out BRG population, both in $M_V$ and $(B-V)_O$, which is
contrary to what is observed. In order to populate the MS up to
observed bright end,
an age 6 Myr less than that obtained from BRGs is needed, in
all the models. This may indicate that the stars which populate the tip of
the MS are 6 Myr younger than the BRGs. Bencivenni et al.
(1991) has assigned an age of 8 Myr to this cluster using model 1
and still find a group of 10 stars lying above the observed
bright end o the MS. They
attribute such an evidence to the occurrence of binary accreting systems.

\subsection{Comparison of observed ILFs with simulated ILFs}
Chiosi et al. (1989) show that
the ILF of main sequence (MS) stars normalized to the
number of evolved stars can be used to differentiate among different
evolutionary scenarios, since it is just the ratio of core H to
He burning lifetimes which is very much affected by the mixing scheme
used. Following Chiosi et al. (1989), we derive the ILF normalized
to the number of evolved stars for these clusters  from the observed data
as well as from synthetic CMDs constructed using different SEMs
and compare them in Fig. 6. for the Salpeter value of mass
function slope (x=1.35) and an assumed binary fraction of 30\%.
The observed ILF depends on the number of evolved stars
present in the sample. In order to
minimize the effect of incompleteness and crowding in the cluster
sample, it is restricted to the outer annular region. It is assumed
that such a sample has the same proportion of stars in all mass ranges.
This assumption is valid except in the case of NGC 1711 where slight
mass segregation among the MS stars has been found (cf. Subramaniam et al.
(1993). Also stochastic nature in the star formation if present
can also affect the number of observed evolved stars.

 Since model 4 is available only for masses $\ge$
2 $M_\odot$ it is not possible to compare the ILFs from this model with the
observed ones, instead we compare the ratio of the core He to H
burning lifetimes.
An inspection of Fig 6. indicates that different models
produces best fits for different clusters.
In order to see the
effects of uncertainties in the value of x and binary fraction (BF)
we have varied
one of these two parameters keeping the other constant. The results
of such an exercise is summarized below:
\begin{description}
\item[(i)] For a given model, the value of BF was fixed at 30\%
and the value of x was changed to get the best fit between
observed and  synthetic ILFs. The values of x for the best fits
are tabulated in Table 5.  This table indicates a general trend
that the model 1 needs steeper whereas model 3 needs flatter
mass function slope. The value most deviating from the Salpeter
value is for NGC 1711 in the case of model 1. Hence, it can be
seen that with values of x not very different from Salpeter's
value, the various models are able to fit the observed ILFs.
\item[(ii)] For a given model, the value of x is fixed at 1.35
(Salpeter's value) and the BF is varied to get the best fit. The
values of BF are varied between 0 to 50\% and the results are
tabulated in Table 6. In some cases, this change in BF could not
bring the synthetic ILFs closer to the observed one. Such cases
are given as blanks in the table. It can be seen that to obtain
the best fit some models require rather extreme values of BF.
\end{description}
Thus it can be seen that the comparison of ILFs does not show a
clear view as to which model is to be preferred.
Since the ILFs reflect the ratio of the H to He-burning
lifetimes, the values of He to H-burning lifetime ratios
for the turn-off masses of the clusters as given by the models
are tabulated in Table 7. For a cluster, these values are very different,
while the turn-off masses from the models are not
very different (see table 3).
One striking point being,
both the classical models give very similar values for the ratio.
Hence, it is difficult to find out the value of the ratio which
is fitting the observation, mainly because of the uncertainty in
the value of the mass function slope. Similar conclusion has been
arrived at by Chiosi et al. (1994) while analyzing the SMC
cluster NGC 330.

\section{Conclusions}
We have compared the CMDs and LFs derived from a homogeneous set
of data comprising of the LMC star clusters NGC
1711, NGC 2004, NGC 2164 and NGC 2214 having turn-off masses in
the range 6-15 M$_\odot$ with the synthetic ones
produced using the SEMs from Castellani et al. (1990), Schaller
et al. (1992) and Bressan et al. (1993). The main results of the
present investigation are:

\begin{description}
\item[1.] Classical models make the clusters younger but the
turn-off masses slightly heavier in comparison to the models
incorporating convective core overshooting (CCO).
\item [2.]Stellar evolutionary models with CCO given by Bressan et al.
(1993) reproduce the observed featured of CMDs best amongst the
models used in the analysis.
\item [3.]The synthetic ILF derived from a model strongly depends on
the value of the mass function slope, and lightly on the binary
fraction whereas the observed ILFs are affected by the
uncertainty in the number of evolved stars.
Hence the comparison of the synthetic with the observed ILFs
does not favour any model specifically. In order to constrain
the models from the comparison of the synthetic ILFs with the
observed ones, reliable estimates of mass function slope,
binary fraction are desired.

\end{description}

\section{Acknowledgements}
We are grateful to E. Brocato
for giving the preliminary version of the computer code used in
this analysis. We thank H.C. Bhatt and D.C.V. Mallik
for critical reading of the manuscript. Useful comments given by
the referee C. Chiosi is gratefully acknowledged.
\section{References}
Aparicio A., Bertelli G., Chiosi C., Nasi E., Vallenari A., 1990, A\&A 240,
262 \\
Balona L.A., Jerzybiewicz M., 1993, MNRAS 260, 782\\
Bencivenni D., Brocato E., Buonanno R., Castellani V., 1991, AJ 102, 137\\
Bertelli G., Bressan A., Chiosi C., Angerer K., 1986a, A\&AS 66, 191\\
Bertelli G., Bressan A., Chiosi C., Angerer K., 1986b, in The Ages of star
clusters,
   ed. F.Caputo, Mem. Soc. Astron. Ital. 57, 427\\
Bertelli G., Betto R., Bressan A., Chiosi C., Nasi E., Vallenari A., 1990,
A\&AS 85, 845 \\
Bhatia R.K., MacGillivray H.T., 1988, A\&A 249, L5\\
Bhatia R.K., Piotto G., 1994, A\&A (in press) \\
Bressan A., Fagotto F., Bertelli G., Chiosi C., 1993, A\&AS 100, 647\\
Brocato E., Buonnano R., Castellani V., Walker A., 1989, ApJS 71, 25 \\
Caldwell J.A.R., Coulson I., 1985, MNRAS 212, 879\\
Cassetella A., Barbero J., Geyer E.H., 1987 ApJS 64, 83\\
Castellani V., Chieffi A., Straniero O., 1990, ApJS 74, 463\\
Chiosi C., Bertelli G., Meylan G., Ortolani S., 1989, A\&A 219, 167\\
Chiosi C., Vallenari A., Bressan A., Deng L., Ortolani S., A\&A 1994
(accepted) \\
Cox A.N., Stewart J.N., 1970a, ApJS 19, 243\\
Cox A.N., Stewart J.N., 1970b, ApJS 19, 261\\
Cox A.N., Tabor J.E., 1976, ApJS 31, 271\\
Elson R.A.W., 1991, ApJS 76, 185\\
Grevesse N., 1991, A\&A 242,488\\
Hannaford P., Lowe R.M., Grevesse N., Noels A., 1992, A\&A 259, 301\\
Huebener W.F., Merts A.L., Magu N.H., Agro M.F., 1977, Los Alamos
Scientific Laboratory Report LA-6760-M\\
Iglesias C.A., Rogers F.J., Wilson B.G., 1992, ApJ 397, 717\\
Johnson H.L., 1966, AR\&AA 4, 193\\
Kubiak M., 1990, Acta Astronomica 40, 349\\
Kurucz R.L., 1979, ApJS 40, 1\\
Kurucz R.L., 1991, in Stellar Atmospheres: Beyond classical
models, NATO ASI Series C, Vol. 341 \\
Lee, M.G., 1992, ApJ 399, L133\\
Lupton R.A., Fall S.M., Freeman K.C., Elson R.A.W., 1989, ApJ 347, 201\\
Maeder A., Meynet G.., 1989, A\&A 210, 155\\
Maeder A., Meynet G.., 1991, A\&AS 89, 451\\
Mateo M., 1988, ApJ 331, 281\\
Mateo M., Hodge P., 1986, ApJ 311, 133\\
Mermilliod J.C., Mayor M., 1989 A\&A 288, 618\\
Robertson J.W., 1974, ApJ 191, 67\\
Rogers F.J., Iglesias C.A., 1992, ApJs 79, 507\\
Sagar R., Pandey A., 1989, A\&AS 79, 407\\
Sagar R., Richtler T., 1991, A\&A 250, 324 (Paper II)\\
Sagar R., Richtler T., de Boer K.S., 1991a, A\&A 249, L5\\
Sagar R., Richtler T., de Boer K.S., 1991b, A\&AS 90, 387 (Paper
I)\\
Salpeter E.E., 1955, ApJ 121, 161\\
Schaller G., Schearer D., Meynet G., Maeder A., 1992, A\&AS 96, 269\\
Seggewis W., Richtler T., 1989, in: Recent Developments of
Magellanic Cloud Research. A European Colloquium, eds., K.S.de
Boer, F.Spite, G.Stansiska, published by Observatoire de Paris, p.45\\
Stothers R.B., Chin C. -w., 1991, ApJ 381, L67\\
Subramaniam A., Sagar R., Bhatt H.C., 1993, A\&A 273, 100 \\
Vallenari A., Chiosi C., Bertelli G., Meylan G., Ortolani S., 1991, A\&AS 87,
517\\
Vandenbergh D.A., 1983, ApJS 40, 1\\
Visvanathan N., 1985, ApJ 288, 182\\
Walker A.R., 1987, MNRAS 225, 627\\
Welch D.L., Mateo M., C\^{o}t\'{e} P., Fischer P., Madore B.F.,
1991 AJ 101, 490\\
Welch D.L., Mclean R.A., Madore B.F., McAlary C.w., 1987, ApJ 321, 162\\
\appendix
\section{Appendix}
The computer code used to construct synthetic colour magnitude diagrams using
SEMs is described below. This method reveals the time scales
involved in various evolutionary phases thereby allowing to compare the
number of stars predicted by an evolutionary model with the number of
stars observed in the respective phases.
We have included the effects due to contact as well as
optical binaries and the photometric errors.  The details of the computer code
can be summarized as follows.
\begin{enumerate} \item
Monte-Carlo method is used to distribute stars randomly in a mass interval
for a given mass function slope. This randomness also accounts for
the stochastic nature of the mass function. The expression for the initial
mass function is given by,
\begin{equation}
dN = AM^{-(1+x) }dM
\end{equation}
where dN is the number of stars in the mass interval dM and x is the
mass function slope, the Salpeter (1955) value being, 1.35. The constant of
proportionality is fixed by the number of post-main sequence stars.
Different values of x has been used to find the best fit of
synthetic ILFs with the observed ILFs.
\item The cluster age obtained by fitting isochrones as described
earlier
is assigned to the masses thus obtained, if the star formation is
assumed to take place instantaneously. It is also possible to give an
age spread to the masses, if we assume a spread in the formation time
of the cluster members. This spread can either be a gaussian with peak
at the cluster age or a step function.
\item The photometric errors present in the observations will produce a
scatter in the CMD of the cluster. It is essential to include the
photometric errors in the synthetic CMD.
These errors are taken from Paper I. The method adopted
for including these errors is adopted from Robertson (1974).
A gaussian distribution has been assumed for the errors with
the standard deviation $\Delta$B and $\Delta$V given by,
$\Delta B = a(B-b)^2$  and $\Delta V = c(V-d)^2$.  The
coefficients a, b, c and d are chosen to fit the observations
where the errors increase with magnitude.
\item Since the width of the MS and the scatter at the tip of the MS are
produced by the presence of the binaries, inclusion of binaries is
very essential in the synthetic CMD. In the galactic open
clusters the binary stars amount up to 30\% of the cluster
population (Mermilliod \& Mayor (1989)).
Apart from the presence of the
actual binaries, contribution from the optical binaries, which
is instrumental is also
substantial because of the high stellar density. Since it is not
clear how much of binaries the LMC clusters contain, we have
assumed a value of 30\% for the overall binary population.
 The procedure works as follows: to each star of mass
$M_1$ we associate a probability of being a member of a binary system
with a companion of mass $M_2$ such that the mass ratio
$M_1/M_2$ is
confined within a given range. We have chosen the range of mass ratio
as 0.75 - 1.25. These two stars are then replaced by a single star
whose flux in various pass bands is given by the sum of the flux of
the component stars.
\end{enumerate}
In general the main sequence is populated randomly according to a
fixed MF slope until the required number of evolved stars are
obtained. The lower limit of the mass is determined by the
observational limit. In general we have adopted the Salpeter value
for the MF slope.
The age is found by fitting the isochrones as explained earlier. The
age spread is not given in general, i.e., an instantaneous star
formation is assumed. The binary stars are
included and the photometric errors are added to the final $M_V$ and
$(B-V)_0$ values.  The LFs are found in an interval of 0.2 mag from
the MS.

\newpage

Figure Captions:\newline
Fig. 1. The photometric differences with Kubiak (1990) and
Elson (1991) are plotted against our data for the four clusters.
The difference $\Delta$ is in the sense literature minus our CCD
values. (a) and (b) give the differences for V and (B$-$V) with
respect to V whereas (e) gives (B$-$V) differences with respect
to (B$-$V) for the clusters NGC 2004, NGC 2164 and NGC 2214,
which are indicated by different symbols. (c), (d) and (f) are
same as above but for the cluster NGC 1711 and two symbols
indicate two sets of data.

Fig. 2. The cluster CMDs after completeness correction and
field star subtraction are shown here. The regions occupied by MS
stars, BRGs and FRGs are also shown.

Fig. 3. The differential luminosity functions (DLFs) of the
four clusters are plotted in solid lines. The DLFs from the raw data
are indicated in dashed lines, whereas the DLFs from the data after
completeness correction (before field star subtraction) is shown
in dotted lines.

Fig. 4. The determination of the age by isochrone fitting
for the four clusters are shown here. The continuous line
represent model 1, dotted line model 2, short dashed line model
3 and long dashed line model 4.

Fig. 5. The synthetic CMDs those match the observation for
x=1.35 and binary fraction of 30\% for
the four clusters are shown here. The models from which the CMDs are
made are also shown.

Fig. 6. The ILFs for the four clusters are plotted here.
The continuous line indicate the observed ILFs. The dotted line
represent the ILFs from model 1, short dashed line from model 2 and long
dashed
line from model 3. The error bars indicate the statistical error at
the magnitudes shown for the
observed ILF. The synthetic ILFs are constructed for x=1.35 and
a binary fraction of 30\%.

\begin{table*}
\caption[]{Statistical results of the photometric differences
$\Delta$ in the sense
literature minus our CCD values. V and (B$-$V) are from our
photometry.  The mean and standard deviation ($\sigma$ ) are
based on N stars. A few points discrepant by more than 3$\sigma$
have been excluded from the analysis.}
\vspace{0.50cm}
\scriptsize
A.Comparison with Kubiak (1991) CCD data in NGC 1711\flushleft
\begin{tabular}{|rcr|rrr|rrr|rcr|rrr|}
\hline
\m{3}{|c|}{V}&\m{3}{c}{$\Delta$V in mag}&\m{3}{c|}{$\Delta$(B$-$V) in mag}
&\m{3}{c|}{(B$-$V)}&\m{3}{c|}{$\Delta$(B$-$V) in mag} \\
\cline{4-9}  \cline{13-15}
\m{3}{|c|}{(mag)}&Mean&$\sigma$&N&Mean&$\sigma$&N&
\m{3}{c|}{(mag)}&Mean&$\sigma$&\m{1}{c|}{N}  \\
\hline
13.0&-&14.0&$-$0.10&0.03&5&$-$0.16&0.07&4&$-$0.22&-&$-$0.10&0.04&0.04&7  \\
14.0&-&14.8&$-$0.20&0.13&4&$-$0.13&0.10&4&$-$0.10&-&0.10&$-$0.10&0.02&5  \\
14.8&-&15.2&0.08&0.02&8&$-$0.52&0.07&8&0.10&-&1.00&$-$0.22&0.07&4  \\
15.2&-&16.1&$-$0.26&0.24&8&$-$0.07&0.00&8&1.00&-&1.70&0.52&0.00&9 \\
\hline
\end{tabular}

\vspace{0.5cm}

B. Comparison with Elson (1991) CCD data in NGC 1711 \flushleft
\begin{tabular}{|rcr|rrr|rrr|rcr|rrr|}
\hline
\m{3}{|c|}{V}&\m{3}{c}{$\Delta$V in
mag}&\m{3}{c|}{$\Delta$(B$-$V) in mag}
&\m{3}{c|}{(B$-$V)}&\m{3}{c|}{$\Delta$(B$-$V) in mag} \\
\cline{4-9}  \cline{13-15}
\m{3}{|c|}{(mag)}&Mean&$\sigma$&N&Mean&$\sigma$&N&
\m{3}{c|}{(mag)}&Mean&$\sigma$&\m{1}{c|}{N}  \\
\hline
13.5&-&15.0&$-$0.03&0.09&10&$-$0.16&0.03&10&$-$0.20&-&0.20&$-$0.19&0.09&69  \\

15.0&-&16.0&$-$0.06&0.05&10&$-$0.17&0.13&10&0.20&-&1.20&$-$0.07&0.62&7\\
16.0&-&16.5&$-$0.11&0.07&9&$-$0.18&0.08&9&1.20&-&1.60&$-$0.14&0.30&12   \\
16.5&-&17.0&0.00&0.15&22&$-$0.18&0.16&19& & & & & & \\
17.0&-&17.5&$-$0.05&0.15&18&$-$0.09&0.12&15& & & & & & \\
17.5&-&18.2&$-$0.06&0.19&21&$-$0.17&0.09&22& & & & & & \\
\hline
\end{tabular}

\vspace{0.5cm}

C. Comparison with Elson (1991) CCD data in NGC 2004 \flushleft
\begin{tabular}{|rcr|rrr|rrr|rcr|rrr|}
\hline
\m{3}{|c|}{V}&\m{3}{c}{$\Delta$V in
mag}&\m{3}{c|}{$\Delta$(B$-$V) in mag}
&\m{3}{c|}{(B$-$V)}&\m{3}{c|}{$\Delta$(B$-$V) in mag} \\
\cline{4-9}  \cline{13-15}
\m{3}{|c|}{(mag)}&Mean&$\sigma$&N&Mean&$\sigma$&N&
\m{3}{c|}{(mag)}&Mean&$\sigma$&\m{1}{c|}{N}  \\
\hline
13.0&-&15.0&$-$0.08&0.09&8&$-$0.01&0.17&9&$-$0.20&-&0.20&$-$0.02&0.09&52  \\
15.0&-&16.0&$-$0.12&0.16&12&0.00&0.07&13&0.20&-&1.70&0.15&0.16&10  \\
16.0&-&17.0&$-$0.03&0.24&18&$-$0.01&0.10&20&&&&&&  \\
17.0&-&17.5&$-$0.18&0.16&9&$-$0.03&0.16&10&&&&&&  \\
17.5&-&18.6&0.01&0.17&9&$-$0.02&0.08&10&&&&& & \\
\hline
\end{tabular}

\vspace{0.5cm}

D. Comparison with Elson (1991) CCD data in NGC 2164 \flushleft
\begin{tabular}{|rcr|rrr|rrr|rcr|rrr|}
\hline
\m{3}{|c|}{V}&\m{3}{c}{$\Delta$V in
mag}&\m{3}{c|}{$\Delta$(B$-$V) in mag}
&\m{3}{c|}{(B$-$V)}&\m{3}{c|}{$\Delta$(B$-$V) in mag} \\
\cline{4-9}  \cline{13-15}
\m{3}{|c|}{(mag)}&Mean&$\sigma$&N&Mean&$\sigma$&N&
\m{3}{c|}{(mag)}&Mean&$\sigma$&N  \\
\hline
13.5&-&15.0&$-$0.03&0.06&7&$-$0.12&0.05&7&$-$0.20&-&0.20&$-$0.10&0.06&21  \\
15.0&-&16.0&$-$0.10&0.08&10&$-$0.11&0.08&9&0.20&-&1.60&$-$0.09&0.14&15 \\
16.0&-&17.0&$-$0.06&0.15&12&$-$0.08&0.14&12&&&&&&  \\
17.0&-&17.6&0.01&0.18&8&$-$0.08&0.08&8&&&&&&  \\
\hline
\end{tabular}
\vspace{0.5cm}

E. Comparison with Elson (1991) CCD data in NGC 2214 \flushleft
\begin{tabular}{|rcr|rrr|rrr|rcr|rrr|}
\hline
\m{3}{|c|}{V}&\m{3}{c}{$\Delta$V in
mag}&\m{3}{c|}{$\Delta$(B$-$V) in mag}
&\m{3}{c|}{(B$-$V)}&\m{3}{c|}{$\Delta$(B$-$V) in mag} \\
\cline{4-9}  \cline{13-15}
\m{3}{|c|}{(mag)}&Mean&$\sigma$&N&Mean&$\sigma$&N&
\m{3}{c|}{(mag)}&Mean&$\sigma$&N  \\
\hline
13.0&-&16.0&$-$0.08&0.06&7&0.03&0.16&8&$-$0.20&-&0.20&$-$0.09&0.12&12  \\
16.0&-&17.0&0.04&0.12&8&$-$0.05&0.13&6&0.20&-&1.70&0.05&0.16&9 \\
17.0&-&18.0&0.02&0.18&8&$-$0.08&0.15&7&&&&&&  \\
\hline
\end{tabular}
\end{table*}
\normalsize
\clearpage
\begin{table*}
\caption[]{Coordinates of cluster centre ($X_c$, $Y_c$), annuli of selected
rings
in pixels and E(B$-$V) values in mag for the clusters are
given.}
\vspace{0.5cm}
\begin{tabular}{|c|cc|cc|c|}
\hline
Cluster & $X_c$ & $Y_c$ &\m{2}{c|}{ Selected annuli}& E(B$-$V) \\ \cline{4-5}
& & & Ring 1 & Ring 2 & \\
\hline
NGC 1711 & 140 & 242 & $50 \leq R<115$ & $115 \leq R<200$ & 0.09\\
NGC 2004 & 150 & 407 & $45 \leq R<115$ & $115 \leq R<200$ &0.09 \\
NGC 2164 & 132 & 237 & $45 \leq R<110$ & $110 \leq R<180$ & 0.10 \\
NGC 2214 & 171 & 231 & $45 \leq R<110$ & $110 \leq R<200$ & 0.07 \\
\hline
\end{tabular}
\end{table*}

\begin{table*}
\caption[]{Age and turn-off mass of the clusters determined from
the models.}
\vspace{0.5cm}
\begin{tabular}{|c|cccc|rrrr|}
\hline
Cluster & \m{4}{c|}{Ages in Myr}& \m{4}{c|}{Turn-off
masses in $M_\odot$}  \\
 & \m{4}{c|}{from models} &
\m{4}{c|}{from models}
\\ \cline{2-9}
& 1 & 2 & 3 & 4 & 1 & 2 & 3 & 4 \\
\hline
NGC 1711 & 22 & 28 & 35 & 23 & 9.7 & 8.3 & 8.3 & 9.4 \\
NGC 2004 & 12 & 14 & 18 & 12 & 14.3 & 12.3 & 11.4 & 13.7 \\
NGC 2164 & 35 & 50 & 60 & 38 & 7.5 & 6.4& 6.4 & 7.2 \\
NGC 2214 & 37 & 50 & 60 & 40 & 7.3 & 6.4 & 6.4 & 7.0 \\
\hline
\end{tabular}
\end{table*}
\begin{table*}
\caption[]{ The observed features of the cluster CMDs are compared with the
ones predicted by the models from the synthetic CMDs. In the table,
(a), (b), (c) and (d) denote the M$_V$ of the MS
tip, M$_V$ of the faintest BRG,
number of blue supergiants and the (B$-$V) value of
the reddest BRG respectively. The last column gives the error in the values
obtained from models.}
\vspace{0.5cm}
\begin{tabular}{|ccrrrrrr|}
\hline
cluster & quantity& observed&\m{5}{c|}{Values from models}\\
 & & & 1 & 2 & 3 & 4 & error  \\
\hline
NGC 1711 & (a) & $-$4.5 & $-$4.0 &$-$3.6 &$-$4.0 & $-$3.9 & $\pm$ 0.2\\
         & (b) & $-$3.8 & $-$4.5 & $-$4.5 & $-$4.4 & $-$4.3 & $\pm$ 0.1\\
         & (c) & 1    & 2   &   4  &   0 & 3 &\\
         & (d) & 1.55& 1.3& 1.7 & 1.65 & 1.34 & $\pm$ 0.2\\
NGC 2004 & (a) & $-$5.5 & $-$4.0 & $-$4.75 & $-$4.6 & $-$4.2& $\pm$ 0.1 \\
         & (b) & $-$4.45 & $-$5.8 & $-$5.6 & $-$5.35 & $-$5.0 & $\pm$ 0.1\\
         & (c) & 0 & 0 & 2 & 0 & 4 & \\
         & (d) & 1.9 & 1.65 & 1.63 & 1.45 & 1.40 & $\pm$ 0.2\\
NGC 2164 & (a) & $-$3.3 & $-$3.0 & $-$3.3 & $-$3.4 & $-$3.6 & $\pm$ 0.3\\
         & (b) & $-$3.0 & $-$3.8 & $-$3.2 & $-$3.1 & $-$3.4 & $\pm$ 0.1\\
         & (c) & 1 & 0 & 6 & 0 & 5 &\\
         & (d) & 1.38 & 1.23 & 1.47 & 1.46 & 1.24 & $\pm$ 0.1\\
NGC 2214 & (a) & $-$3.6 & $-$3.3 & $-$3.3 & $-$3.4 & $-$3.1 & $\pm$ 0.3\\
         & (b) & $-$3.2 & $-$3.5 & $-$3.5 & $-$3.1 & $-$3.3 & $\pm$ 0.1\\
         & (c) & 0 & 2 & 4 & 0 &4 & \\
         & (d) & 1.3 & 1.0 & 1.48 & 1.2 & 1.25 & $\pm$ 0.1\\
\hline
\end{tabular}
\end {table*}

\begin{table}
\caption[]{The values of x which produces the best fit between
synthetic and observed ILFs for a binary fraction of 30\% for
different models.}
\vspace{0.5cm}
\begin{tabular}{|r|lll|}
\hline
cluster&\m{3}{c|}{x value for models} \\
   & 1& 2 & 3 \\ \hline
NGC 1711 & 1.90 & 1.60 & 1.35 \\
NGC 2004 & 1.65 & 1.35 & 1.25 \\
NGC 2164 & 1.35 & 1.13 & 1.05 \\
NGC 2214 & 1.35 & 1.25 & 1.10 \\
\hline
\end{tabular}
\end{table}

\begin{table}
\caption[]{The value of binary fraction which produces the best
fit between the observed and synthetic ILFs for the Salpeter
mass function slope (x=1.35) for different models.}
\begin{tabular}{|r|lll|}
\hline
Cluster& \m{3}{c|}{value of the BF for models} \\
  & 1 & 2& 3 \\
\hline
NGC 1711 & -- & 50\% & 30\% \\
NGC 2004 & 50\% & 30\%& 0\%\\
NGC 2164 & 30\%& 20\%& --\\
NGC 2214 & 30\%& 20\%& 5\% \\
\hline
\end{tabular}
\end{table}

\begin{table*}
\caption[]{The values for the ratio of the helium
to hydrogen burning lifetime is listed for different masses.
The masses are in solar units.}
\vspace{0.5cm}
\begin{tabular}{|r|llll|}
\hline
Mass&\m{4}{c|}{$\tau_{He}/\tau_H$} \\ \cline{2-5}
& Model 1 & Model 2 & Model 3& Model 4\\
\hline
7.0 & 0.169 & 0.109 & 0.071 & 0.146 \\
9.0 & 0.144 & 0.099 & 0.063 & 0.111 \\
12.0 &      & 0.098 & 0.062 & 0.087 \\
15.0 &      & 0.096 & 0.065 & 0.093 \\
\hline
\end{tabular}
\end{table*}

\end{document}